# Trapping of isotropic droplets by disclinations in nematic liquid crystals controlled by surface anchoring and elastic constant disparity.

Nilanthi P. Haputhanthrige[1,2], Sathyanarayana Paladugu[1], Maxim O. Lavrentovich[3,4], Oleg D. Lavrentovich[1,2,5*]

[1] Advanced Materials and Liquid Crystal Institute, Kent State University, Kent, OH 44242, USA
[2] Department of Physics, Kent State University, Kent, OH 44242, USA
[3] Department of Earth, Environment, and Physics, Worcester State University; Worcester, MA 01602, USA
[4] Department of Physics and Astronomy, University of Tennessee; Knoxville, Tennessee 37996, USA
[5] Materials Science Graduate Program, Kent State University, Kent, OH 44242, USA

**Abstract.** Linear defects such as dislocations and disclinations in ordered materials attract foreign particles since they replace strong elastic distortions at the defect cores. In this work, we explore the behavior of isotropic droplets nucleating at singular disclinations in a nematic liquid crystal, predesigned by surface photopatterning. Experiments show that in the biphasic nematic-isotropic region, although the droplets are attracted to the disclination cores, their centers of mass shift away from the core centers as the temperature increases. The shift is not random, being deterministically defined by the surrounding director field. The effect is explained by the balance of interfacial anchoring and bulk elasticity. An agreement with the experiment can be achieved only if the model accounts for the disparity of the nematic elastic constants; the so-called one-constant approximation, often used in the theoretical analysis of liquid crystals, produces qualitatively wrong predictions. In particular, the experimentally observed shift towards the bend region around a +1/2 disclination core can be explained only when the bend constant is larger than the splay constant. The described dependence of the precise location of a foreign inclusion at defect cores on the elastic and surface anchoring properties can be used in rational design of microscale architectures.

# I. INTRODUCTION

Topological defects play an important role in the behavior of liquid crystals. George Friedel coined the term "nematic" for the simplest type of a liquid crystal while observing flexible defect lines, known nowadays as disclinations, in specimens of uniaxial paraelectric nematic (N) [1]. Topologically stable disclinations are the ones around which the director $\hat{n}$ of molecular orientation rotates by 180°; these are assigned a topological charge or strength +1/2 or -1/2, depending on the sense of director rotation in relation to the direction of circumnavigation around the defect [2,3]. The director experiences strong gradients at disclination cores. The rapid increase of the gradient energy near the cores controls the spatial distribution of chemical species: foreign particles are attracted to the cores since they replace the director gradients and reduce the associated large core energy. This effect has been observed in different settings, such as the deposition of polymer [4] and surfactant [5,6] molecules at the cores and attraction of colloidal spheres [7-12] and rods [13-16] to defect cores. The recent development of patterning techniques to create disclination networks [17-27] spurred the exploration of predesigned disclinations to control the spatial distribution of inanimate [27-32] and living microparticles [33,34].

An important facet of particle-disclination interaction is the nematic-to-isotropic (NI) phase transition. Due to its first order character, the transition proceeds through nucleation of isotropic droplets that expand and replace the N phase. There is a finite temperature range over which the two phases coexist [2]. Nucleation of the isotropic phase is facilitated if the sample contains inhomogeneities such as dust particles and disclinations. Evidence of such heterogeneous nucleation is presented in Fig.1, which shows an NI coexistence in an unaligned sample. The sample features topological defects, including singular disclinations of strength $\pm 1/2$ emitting two brushes of extinction from their cores and surface point defects-boojums [35] producing four extinction brushes. As the sample is heated, the isotropic islands appear at the $\pm 1/2$ disclination cores (marked by red arrows), boojum cores (black arrows), and also in the defect-free areas, which likely contain microscopic impurities (yellow arrows).

Mottram and Hogan [36] and later Mottram and Sluckin [37] developed a theoretical model that justified nucleation of isotropic droplets at the disclinations and predicted that the NI transition at the cores could start at temperatures below the temperature of the NI transition in a homogeneous, uniform N. Surface photopatterning allows us to explore the roles of defects in phase transitions by controlling the position and character of the disclinations in a liquid crystal. As demonstrated by Chen et al. [27] for thin layers of a single-compound N pentylcyanobiphenyl (5CB), the predesigned topological defects do indeed serve as nucleation sites of the NI transition, although the $\pm 1/2$ cores produce isotropic islands mostly when the heating rate is high, 0.5 °C/min; when the rate is low, 0.02 °C/min, the isotropic phase emerges at random locations, including away from the defect cores. Repeated heating-cooling cycling reveals that the isotropic droplets appear at the same locations outside the cores, indicating that nucleation is caused by dust particles or surface imperfections [27]. Most recently, Han et al. [38] explored nematic shells, in which the disclinations of strength +1/2 are necessitated by the Poincaré–Hopf topological theorem; nucleation of the I phase started predominantly at the defect cores.

In this paper, we explore the NI transition scenario in a thin layer of an N mixture E7 with an array of $\pm 1/2$ disclinations fabricated by photopatterning [18,19,26,27]. The experiments performed at a low heating rate of 0.02 °C/min demonstrate that the disclination cores indeed cause heterogeneous nucleation of the isotropic phase; the isotropic droplets at the cores coexist with the surrounding N background. Also, nucleation at the cores is only observed in cells that underwent a substantial cleaning, which reduces the number of heterogeneous nucleation sites outside the cores. An intriguing feature of the isotropic island growth during heating is that the islands shift away from the disclination cores. These shifts are not random: The islands at the cores of comet-like +1/2 defects always shift along the symmetry axis of the defect, away from its "tail" (where the director is perpendicular to the NI interface) and toward the "head" (where the director is tangential to the interface). We propose theoretical models that explain the shifts by: (a) elastic anisotropy of the N phase, namely, disparity of the elastic constants, and (b) anisotropy of surface tension, known as surface anchoring of the director [2,3], at the NI interface. In the experimentally explored E7 case, the two factors compete with each other; when the elastic anisotropy, caused by a stronger bend modulus as compared to the splay modulus, outweighs the surface anchoring contribution, the model predicts shifts from the tail to the head, as experimentally observed. The deterministic direction and shift magnitude of the foreign inclusions from the disclination cores can be used to precisely control the microscale assembly of particles by disclination networks.

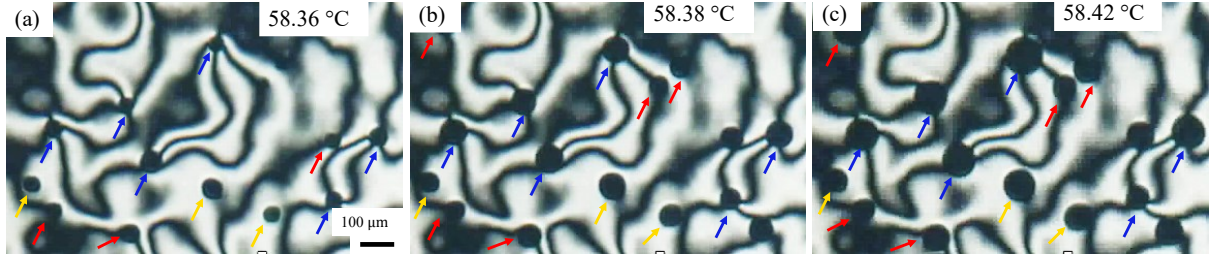

**Figure 1. Polarized optical microscope textures of a progressing NI phase transition in a thin slab of E7 between two glass substrates.** (a-c) textural development with the isotropic nuclei (black inclusions) appearing either at the disclination cores (marked with red arrow), boojum cores (blue arrow), or outside the cores (yellow arrow) at 58.36 °C, 58.38 °C, and 58.42 °C respectively. The cell thickness is $(2.4 \pm 0.1)$ μm.

## II. MATERIALS AND METHODS

Indium tin oxide (ITO) coated glass plates of 1.1 mm thickness are used as substrates. The ITO-coated glass plates are ultrasonically washed in water with a small amount of detergent at 60 °C for 15 min; if the washing time is less than 15 min, nucleation would often start at random locations. The plates are then washed in isopropanol, baked in an oven at 80 °C for 15 min, and exposed to UV in an ozone chamber for 15 min on the ITO-coated side. Once cleaned, the ITO-coated glass plates are spin-coated with a photosensitive alignment agent SD1 dissolved in N, N-dimethylformamide (DMF) at a concentration 0.5 wt. % [39] at a spin speed of 3000 rpm for 30 s, then baked on a hot plate at 95 °C for 30 min to remove the solvents. Two SD1-coated glass substrates are assembled to form a cell with the SD1 coatings facing inward. Fiber spacers are used to control the cell thickness $h$ in the range 1.2-3.3 μm.

A plasmonic photopatterning technique detailed in [18,19,26,27] is used to pattern the director field. First, one designs and fabricates a plasmonic metamask with the desired geometrical pattern of rectangular nanoapertures. The nanoapertures are positioned in a triangular lattice with a periodicity of 270 nm. A light beam passing through the mask acquires local linear polarization along the short axis of the nanoapertures. The transmitted light beam with patterned polarization irradiates an empty cell made of two SD1-coated glass plates, producing identical patterns of SD1 molecular orientation at them. The elongated SD1 molecules align perpendicularly to the local polarization of light. As a result, the imprinted molecular orientation of SD1 replicates the pattern of nanoslits. The cells are then filled with a nematic mixture E7 through capillary suction at a temperature above the NI transition temperature. The cell is cooled down to the N phase. The SD1 films transfer the imprinted molecular orientation onto the nematic director, which is tangential to the substrates.

The temperature of cells is controlled by an Instec temperature controller mk2000 with a hot stage HCS 402. The temperature stabilization is better than 0.01 °C. A polarizing optical microscope (Olympus BX51) is used to observe the textures. An EXICOR PolScope MicroImager (Hinds Instruments) maps two-dimensional (2D) patterns of the director and optical retardance $\varGamma$; the isotropic islands are readily recognized as regions with a vanishing $\varGamma$ [40]. The typical phase diagram is N → 58.4 °C → NI → 59.5 °C → I. The temperatures of the transition points might vary by about 1 °C in different cells and for different batches of E7. The extension of the NI biphasic region is in the range $1.0 - 1.7$ °C. The biphasic range in single-compound materials such as 5CB is more narrow, less than 1 °C, which makes the analysis of the nuclei more difficult; however, the behavior of the islands was qualitatively similar to the one in the E7 mixture. We use a relative temperature, $T - T_{NI} > 0$, where $T_{NI}$ is the nematic-to-biphasic transition point. The areas and coordinates of the centers of mass of the I islands are measured using the FIJI ImageJ software.

An important parameter of the NI interface is the direction of the "easy" axis, which is the angle $\theta_{eq}$ that the equilibrium director $\hat{\mathbf{n}}_{eq}$ makes with the normal $\hat{\mathbf{v}}$ to the interface. For 5CB, Faetti and Palleschi [41] determined that this angle is $\theta_{eq} = (63 \pm 6)°$. To estimate $\theta_{eq}$ in E7, we use glass plates with spin-coated polystyrene (PS), 0.5 wt% solution in chloroform. PS sets a degenerate in-plane orientation of cyanobiphenyls [42]. A small temperature gradient is created by inserting a 20 μm thick insulator between the cell and the hot stage at one edge of the cell. The temperature gradient results in an extended NI interface. The PolScope MicroImager's map of the director, Fig. 11 in Appendix A, suggests that $\hat{\mathbf{n}}_{eq}$ is tilted by $\theta_{eq} = (61 \pm 8)°$ from the normal to the interface.

We consider director patterns with either isolated +1/2 and -1/2 defects, or periodic arrays of +1/2 and -1/2 defects, written in the Cartesian coordinates $(x, y, z)$ as $\hat{\mathbf{n}}_0 =$

$$(n_x, n_y, 0) = (\cos\varphi, \sin\varphi, 0) \,, \quad \text{where } \varphi = \frac{1}{2}\sum_{i=1,j=1}^{i=p,j=q}\left[(-1)^{i+j}\arctan\left(\frac{y-jb}{x-ia}\right) + \pi\sin^2\left(\frac{\pi j}{2}\right)\right],$$

where $p$ and $q$ are the numbers of defects in rows and columns, respectively. $a$ and $b$ are the distances between the cores along the $x$ and $y$ directions, respectively. We choose $p = q = 14$ and $a = b = 98$ μm. The phase $\frac{\pi}{2}\sin^2\left(\frac{\pi j}{2}\right)$ adopts a value of 0 or $\pi/2$, defining the left-right orientation of the defects. The pattern is written as the superposition of the director fields of individual $\pm 1/2$ defects, which is justified in the one-constant approximation [2]. Optical microscopic observations with two crossed polarizers reveal Schlieren textures with two black brushes emanating from each defect core, indicating that the topological charges of these defects are either $+1/2$ or $-1/2$, Fig.2a. These observations, however, do not lift the well-known degeneracy of the two mutually orthogonal director fields [3], which could be associated with the textures such as the ones in Fig.2a. To lift this degeneracy, one can insert an optical compensator with its slow axis bisecting the directions of polarizer and analyzer, Fig.2b. However, this mode of observation still provides only a qualitative map of the director. PolScope Microimager observations, Fig.2c provide a quantitative mapping. The PolScope is a variable optical compensator that records the image of a birefringent sample for several preselected retardance values. These patterns allow one to extract two separate maps: the director field projected onto the plane of the sample, shown by ticks in Fig.2c, and the optical retardation, represented by pseudocolors in Fig.2c. The optical retardation is independent of the in-plane orientation of the director, provided its component along the direction of light propagation does not change within the explored area. The PolScope mode makes it easy to locate and characterize the I nuclei since these regions are of zero birefringence and appear dark in the pseudocolor chart of the PolScope scale, Fig.2c.

    The surface anchoring patterns are the same on the top and bottom plates. The analyzed nuclei are wider than the thickness of the cell, which is in the range from 1.2 to 3.2 μm, and thus contact both substrates. Smaller nuclei might touch only one substrate or levitate in the bulk [43], but we do not discuss these situations. Thus, the system in the nematic state can be considered as approximately two-dimensional, with singular $\pm 1/2$ disclination lines connecting the top and bottom surfaces. As shown in Fig. 12 in Appendix A, different wettability of the glass substrates by the N and I phases produces a meniscus effect; for thin cells, this effect is small, as the variation of the NI interface along the normal z-axis is 1

µm or less. The effect of the meniscus on the behavior of the isotropic droplets is discussed in more detail in Appendix B.

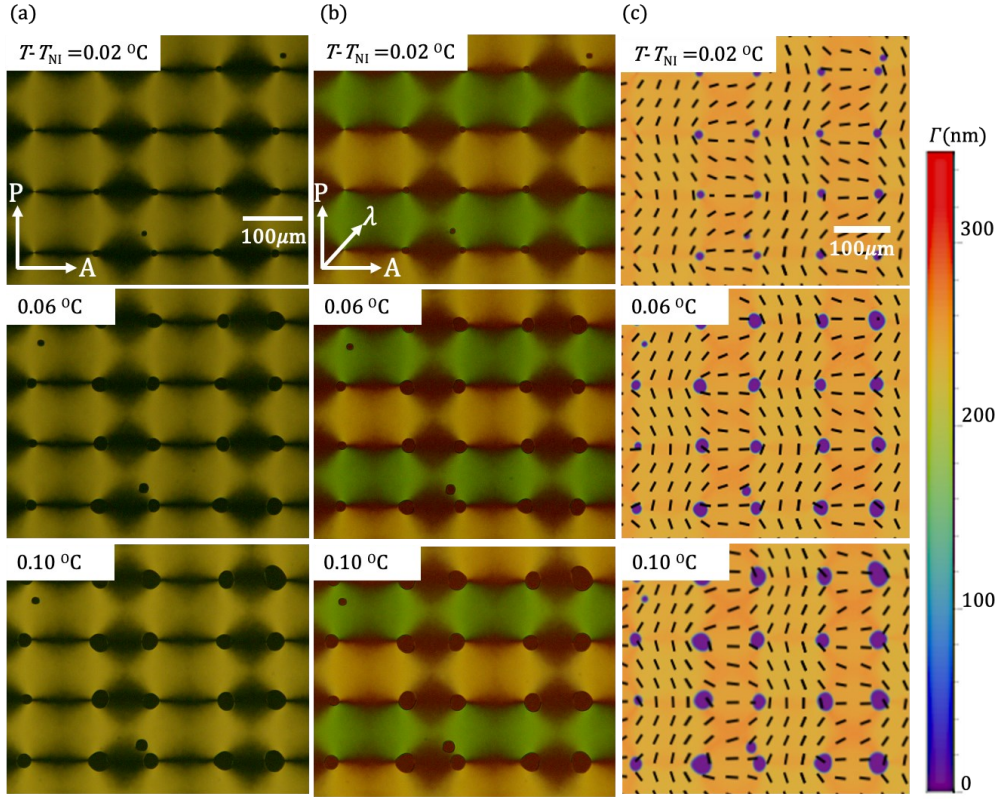

**Figure 2. The progressive growth of I islands in an array of $\pm 1/2$ defects**; The textures captured using a polarized optical microscope (a) with crossed polarizer and (b) with crossed polarizers and full-wavelength optical compensator, and (c) The textures captured using Polscope MicroImager. P, A, and $\lambda$ represent the polarizer analyzer and the slow axis of the full-wavelength compensator, respectively. Cell thickness $h = (2.3 \pm 0.1)$ µm.

## III. EXPERIMENTAL RESULTS

### A. Isolated $\pm 1/2$ disclinations

The isolated +1/2 and -1/2 disclinations are patterned over a square area 1 mm × 1 mm, with the core at the center of the square. In order to minimize the effect of lateral edges on the behavior of the I islands, the island growth is monitored only until the islands reach a radius of $R \approx 25$ µm. As the temperature increases, the I phase nucleates and grows at the disclination core, Fig.3a. Other nucleation sites are also observed, presumably associated with surface and bulk irregularities. To stabilize the I droplets nucleating at the disclination cores, the temperature is cycled as follows: heating by 0.02 °C/min is followed by a 0.01 °C/min cooling, repeated for 15 to 20 cycles. The nematic director field mapped by the PolScope MicroImager around the I islands, Fig.3a,b, does not show noticeable changes when the nuclei expand, which suggests that the azimuthal surface anchoring at the glass plates [44] is sufficiently strong in the N phase. The area $A$ of the I inclusions increases with temperature, Fig.3d. A notable feature is that the "center of mass" of the I droplets shifts away from the cores of both the +1/2, Fig.3a, and the -1/2 disclinations, Fig.3b. The shift $d_x^{+1/2}$ is along the symmetry $x$-axis of the +1/2 defect, towards the bend region and tangential director alignment at the NI interface, i.e., $d_x^{+1/2} < 0$; $|d_x^{+1/2}|$ can reach 4 μm, Fig.3a. The shift is also quantified by a dimensionless ratio $d_x^{+1/2}/R$, where $R$ is the effective radius of the island, calculated as $R = \sqrt{A/\pi}$, Fig.3d.

The I nuclei at the -1/2 disclination cores show similar behavior, growing with the temperature and shifting away from the cores towards the side where the director is perpendicular to the NI interface, $d_x^{-1/2} > 0$, Fig.3b; the shift is noticeably smaller (by a factor of 3-4) than in the +1/2 core case.

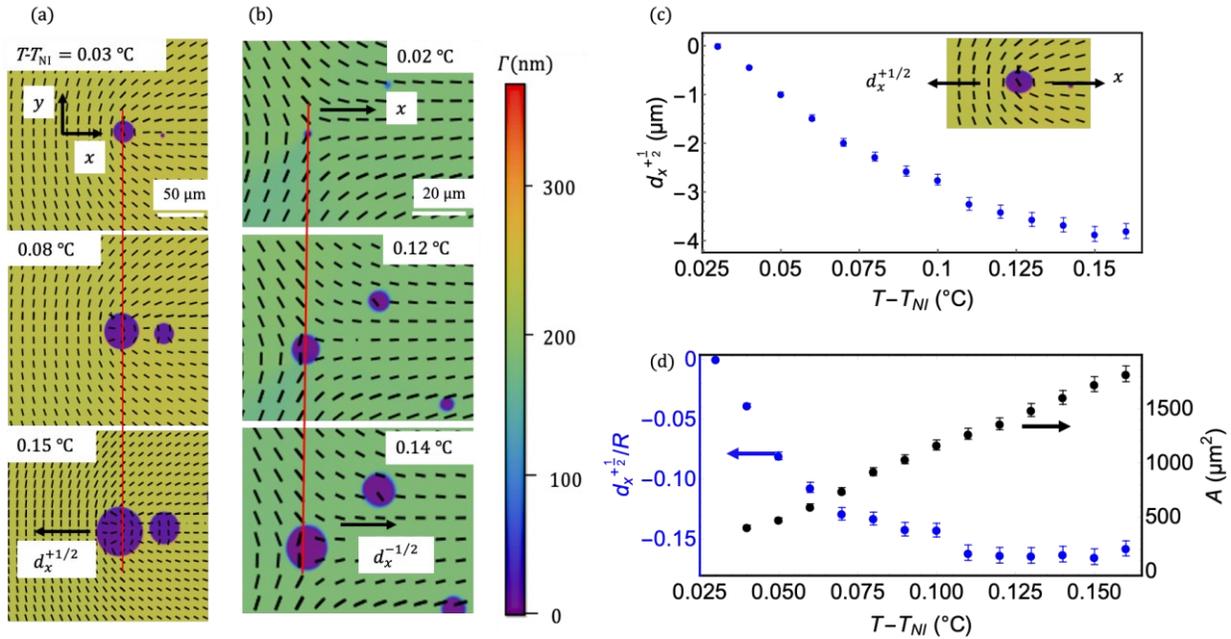

**Figure 3. Temperature dependencies of the shift of isotropic nuclei at the ±1/2 disclination cores:** (a) PolScope Microimager textures of an I nucleus at the +1/2 core; note the shift $d_x^{+1/2}$ towards the tangential anchoring; (b) textures of the I nucleus at the −1/2 core; the shift $d_x^{-1/2}$ is towards the region of homeotropic anchoring; (c) temperature

dependence of $d_x^{+1/2}$; (d) temperature dependence of the normalized shift $d_x^{+1/2}/R$ and the area $A$ of the growing I island; the effective radius of the island is $R = \sqrt{A/\pi}$. Cell thickness is $h = (3.2 \pm 0.1)$ μm in (a,c,d) and $(1.7 \pm 0.1)$ μm in (b).

## B. Square arrays of $\pm 1/2$ disclinations

To establish the island shift statistics, we explore square arrays of alternating +1/2 and -1/2 disclinations. Nucleation starts at the defect cores when the cells are prepared from well-cleaned plates and heated slowly. The I islands grow and shift away from the cores, Fig.4a,b. Similarly to the case of isolated $\pm 1/2$ disclinations, Fig.3, the droplets at +1/2 cores shift towards the tangential anchoring, while the droplets at -1/2 cores shift towards the homeotropic anchoring along the symmetry axis, Fig. 4a. The shifts $|d_x^{+1/2}|$ are about 3-4 times larger than the shift $|d_x^{-1/2}|$, Fig.4,c. For example, at $\Delta T = 0.14$ °C, $d_x^{+1/2} \approx -4$ μm, while $d_x^{-1/2} \approx 1$ μm, Fig.4(b).

In the arrays, each +1/2 defect has two -1/2 neighbors along the $x$-axis of the islands' shift, separated in the N state ($\Delta T = 0$) by a distance $D_{\parallel 0} = D_{\perp 0} = (98 \pm 2)$ μm. The subscript "$\parallel$" means that the director $\hat{n}_0$ in the region between two defects is parallel to the line that connects them, while "$\perp$" means that $\hat{n}_0$ is perpendicular to the connecting line at the midpoint. In the latter "$\perp$" case, the +1/2 and -1/2 defects share a bend region. Thus, the data on isolated disclinations in Fig.3 suggest that the corresponding separation $D_\perp = D_{\perp 0} - |d_x^{+1/2}| + |d_x^{-1/2}|$ should decrease as the temperature increases by $-|d_x^{+1/2}| + |d_x^{-1/2}| \approx -3$ μm;. For the same reason, the distance from a +1/2 island to the -1/2 island in the "$\parallel$" case should increase, $D_\parallel = D_{\parallel 0} + |d_x^{+1/2}| - |d_x^{-1/2}|$, by about $|d_x^{+1/2}| - |d_x^{-1/2}| \approx 3$ μm; here $D_{\parallel 0} = D_{\perp 0} = (98 \pm 2)$ μm. The temperature dependencies of $D_\perp$, Fig.4d and $D_\parallel$, Fig.4e, confirm these trends, but not exactly. Namely, $D_\parallel$ increases slightly less than expected: $D_\parallel \approx D_{\parallel 0} + 2$ μm, while $D_\perp$ shrinks slightly more than expected, $D_\perp \approx D_{\perp 0} - 5$ μm, so that $\frac{D_{\parallel 0} + D_{\perp 0}}{2} - \frac{D_\parallel + D_\perp}{2} \approx 1.5$ μm rather than zero. This discrepancy, although slight as compared to the initial separations $D_{\parallel 0} = D_{\perp 0} = (98 \pm 2)$ μm, signals that the shifts are affected by a weak elastic attraction of the islands residing at the cores of the opposite strength. The potential of attraction in a one-constant approximation for splay and bend moduli is [3] $f_{att} \approx \frac{\pi}{2} K \ln \frac{D_{\parallel,\perp}}{2R}$ per unit length along the $z$-axis normal to the cell, with $K$ the average elastic constant. Nevertheless, the fact that the shifts at the isolated defects, Fig.3, and in the arrays, Fig.4, are similar and that $D_\parallel > D_{\parallel 0}$ in Fig.4e (which means that the +1/2 and -1/2 droplets *move away from each other*) suggests that the elastic interactions cannot be the principal reason for the shifts.

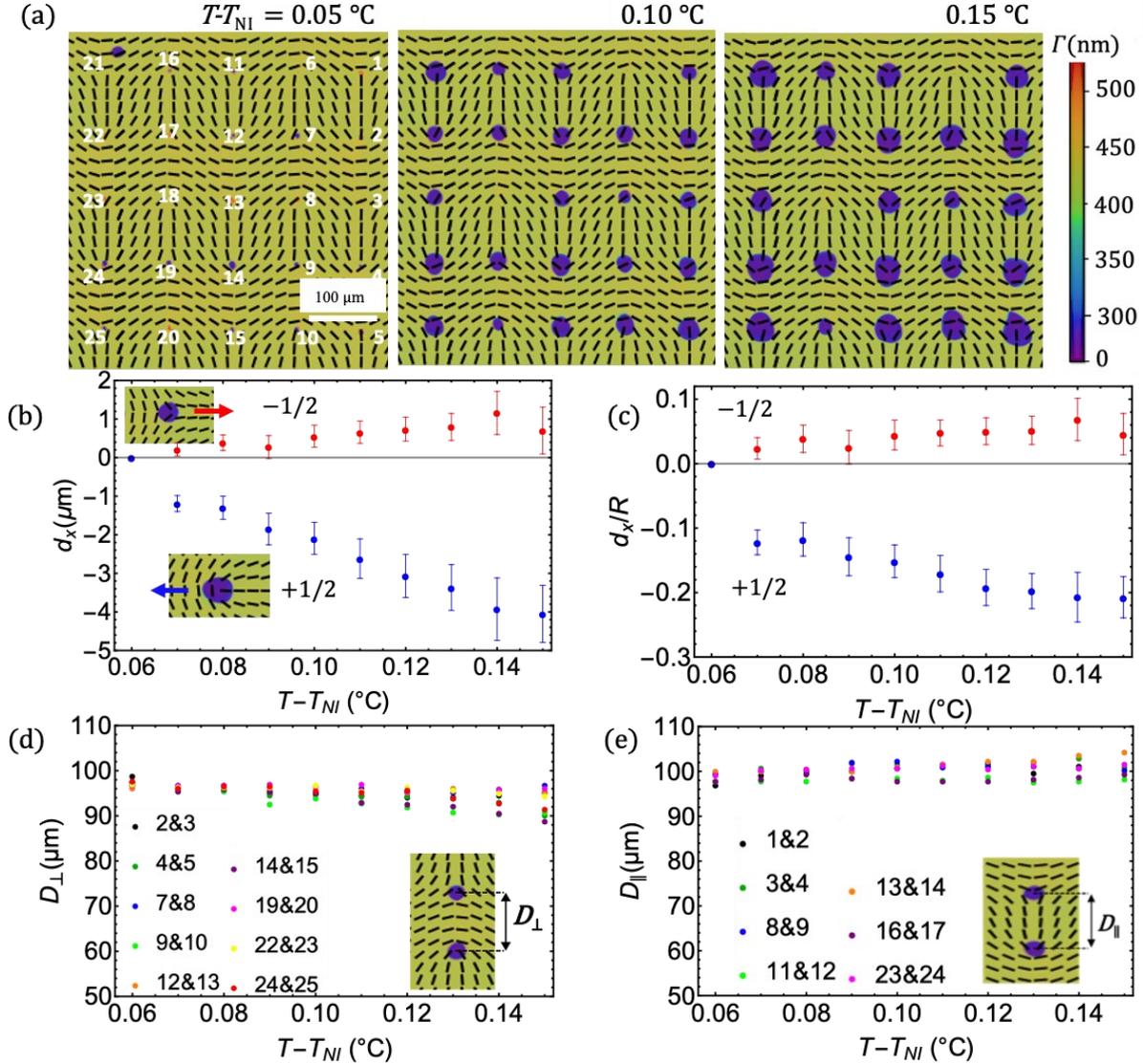

**Figure 4. The shift of isotropic nuclei away from +1/2 and -1/2 cores in the biphasic region of E7.** (a) Textures of the progressive growth of the I islands; (b) absolute and (c) normalized shifts of the islands from -1/2 and +1/2 cores vs. temperature; (d) and (e) distances $D_\perp$ and $D_\parallel$ between two neighboring I islands vs. temperature separated by the bend and splay regions, respectively. Cell thickness $h = 3.25 \pm 0.02$ μm.

Elastic interactions of droplets with the surrounding director field of an isolated disclination also cannot explain the observed shifts if the consideration is limited by a one-constant approximation, in which the Frank-Oseen free energy density writes $\frac{K}{2}[(\operatorname{div}\hat{\mathbf{n}})^2 + (\hat{\mathbf{n}} \times \operatorname{curl}\hat{\mathbf{n}})^2]$. Placement of a circular I island of a radius $R$ centered at the core of an isolated $\pm 1/2$ disclination yields an elastic energy gain $f_e = \frac{\pi K}{4}\ln\frac{R}{r_c}$ since the island removes the director gradients; here $r_c$ is the radius of an unoccupied defect core. This gain is somewhat larger than the attraction elastic energy in the arrays, $\frac{f_e}{f_{att}} = \frac{\ln\frac{R}{r_c}}{2\ln\frac{D_{\parallel,\perp}}{2R}} = 2.1$, if one estimates $D_{\parallel,\perp} = 100$ μm, $R = 20$ μm, and $\frac{R}{r_c} = 10^3$. For a very small island, $R = 3$ μm, the energy gain and the elastic attraction energy are about the same. When the center of the isotropic island shifts away from the core by $\mathbf{d} = (d_x, d_y)$, this elastic gain always decreases, regardless of the azimuthal direction of the shift:

$$f_e = \frac{K}{8}\int_0^{2\pi} d\beta \int_{r_c}^{\rho'} \frac{d\rho}{\rho} = \frac{\pi}{4}K\ln\frac{\sqrt{R^2-d^2}}{r_c}, \qquad (1)$$

where $\rho' = d\cos\beta + d\sqrt{\cos^2\beta - 1 + \frac{r^2}{d^2}}$, $\beta$ is the azimuthal angle of the cylindrical coordinate system, and $d < R$ is the shift length; the sign of the topological charge does not affect the result. The result is clear: the isotropic droplet replaces the high energy of director distortions around the core; any shift of the droplet's center from the core would diminish the energy gain because the droplet would replace a less distorted region. Therefore, the elasticity of the nematic, treated in a one-constant approximation, cannot explain the observations in Figs.3,4. In what follows, we consider two other plausible reasons for the shifts, one associated with the surface anchoring and the second with the inequality of the elastic constants.

## IV. MODEL OF SURFACE ANCHORING EFFECT

A potential reason for the shift of the isotropic islands away from the cores of defects is the anisotropic nature of the molecular interactions at the NI interface, which is reflected in the dependence of the NI interfacial energy on the polar angle $\theta$ of the LC molecules at the interface, measured between the director and the normal $\hat{\mathbf{v}}$ to the interface. At the interface of an apolar nematic and an isotropic fluid, the only angular dependence that could enter the surface energy density is via a term proportional to $(\hat{\mathbf{n}} \cdot \hat{\mathbf{v}})^2$. For the anisotropic surface tension $\sigma_{NI}$ to yield a minimum at some "easy cone" $0 \leq \theta_{eq} \leq \pi/2$, the term $(\hat{\mathbf{n}} \cdot \hat{\mathbf{v}})^2$ must be negative, necessitating higher-order terms in the energy, e.g., $\sigma_{NI} = \overline{W}(\hat{\mathbf{n}} \cdot \hat{\mathbf{v}})^2 + \frac{W}{2}(\hat{\mathbf{n}} \cdot \hat{\mathbf{v}})^4 + const$ (with $\overline{W} < 0$), which can be rewritten as

$$\sigma_{NI} = \sigma_0 + \frac{W}{2}\left[(\hat{\mathbf{n}} \cdot \hat{\mathbf{v}})^2 - (\hat{\mathbf{n}}_{eq} \cdot \hat{\mathbf{v}})^2\right]^2, \quad (2)$$

where $\hat{\mathbf{n}}_{eq}$ is the "easy axis" making the equilibrium angle $\theta_{eq}$ with $\hat{\mathbf{v}}$, defined via $\cos^2\theta_{eq} = -\frac{\overline{W}}{W}$. In 5CB [41], $\theta_{eq} = (63.5 \pm 6)°$, the interfacial tension $\sigma_0 = (1.5 \pm 0.6) \times 10^{-5}$ J/m$^2$, while the anchoring coefficient defined from the potential $\sigma_0 + w_F(\theta - \theta_{eq})^2$ is $w_F = (4.8 \pm 1.2) \times 10^{-7}$ J/m$^2$ rad$^2$, which leads to the estimate of $W$ in Eq.(2): $W = (1.5 \pm 0.4) \times 10^{-6}$ J/m$^2$. The ratio of the anchoring coefficient to the surface tension coefficient, $W/\sigma_0$, strongly affects the properties of the nuclei in NI phase transitions, as discussed by Kim et al. [45]. In lyotropic chromonic nematics, for example, the surface tension anisotropy is very strong, $W/\sigma_0 \sim 1 \div 10$, which leads to tactoidal shapes of the nuclei with cusps at which the N-I interface changes its direction abruptly; the cusps are located in the regions where $\hat{\mathbf{n}}$ is perpendicular to the interface. In our case, the anisotropic part of surface tension is weak, $W/\sigma_0 \sim 0.1$, which explains why the shape of the nuclei is close to circular, Fig.3a,4a, with smooth corners. The corners form at regions where $\hat{\mathbf{n}}$ is tangential to the NI interface.

In a 2D model that assumes (i) no $z$-dependence of the NI interface and (ii) that the director field around circular isotropic islands is fixed by the surface patterning, the shifts $\mathbf{d} = (d_x, d_y)$ might produce a better fit of the director orientation at the NI interface to the easy directions $\theta_{eq}$, and thus a smaller interfacial anchoring energy

$$f_a = \frac{WR}{2}\oint\left[(\hat{\mathbf{n}} \cdot \hat{\mathbf{v}})^2 - \cos^2\theta_{eq}\right]^2 d\tau. \quad (3)$$

The integral in Eq. (3) is calculated over a circle of a radius $R$; $\tau$ is the angular parameter along the circle. We calculate $f_a$ for islands located at isolated defects, with a director field given by

$$\hat{\mathbf{n}} = \left\{\cos\left[m\arctan\left(\frac{y+d_y}{x+d_x}\right)\right], \sin\left[m\arctan\left(\frac{y+d_y}{x+d_x}\right)\right]\right\}, \quad (4)$$

where $m = \pm 1/2$ and the Cartesian coordinates are parametrized by the polar coordinates $(R, \tau)$: $x = R\cos\tau, y = R\sin\tau$, while the normal to the NI interface is

$$\hat{\mathbf{v}} = \{\cos\tau, \sin\tau\}. \quad (5)$$

Below we discuss the droplet at an isolated $m = +1/2$ disclination for four types of NI anchoring: (i) homeotropic, $\theta_{eq} = 0$; (ii) tangential, $\theta_{eq} = 90°$; (iii) intermediate $\theta_{eq} = 45°$, and (iv) $\theta_{eq} = 61°$, estimated for E7 (see Appendix A), and close to the available data on 5CB.

## A. Isolated +1/2 disclination

(i,ii) *Tangential and homeotropic anchoring.* The normalized anchoring energy $f_a/RW$ calculation predicts that the centers of I droplets shift from the defect cores $(x, y) = (0,0)$, Fig.5a. The minimum $\frac{f_a}{RW} = 0.81$ is achieved at $\left(\frac{d_x^{+1/2}}{R}, \frac{d_y^{+1/2}}{R}\right) = (0.91, 0)$ for $\theta_{eq} = \pi/2$, and at $\left(\frac{d_x^{+1/2}}{R}, \frac{d_y^{+1/2}}{R}\right) = (-0.91, 0)$ for $\theta_{eq} = 0$. The predicted shifts are very significant, as the I islands almost leave the core of the defect, Fig.5b. The shifts, as already stated, are opposed by the elastic energy $f_e$. For elastic/anchoring ratio $\frac{\pi K}{4WR} = 0.1$, which would be the case if $K = 2.1$ pN [41], $R = 11$ μm, and $W = 1.5 \times 10^{-6}$ J/m², the shifts are reduced to $\left(\frac{d_x^{+1/2}}{R}, \frac{d_y^{+1/2}}{R}\right) = (0.77, 0)$ for $\theta_{eq} = \pi/2$, and $\left(\frac{d_x^{+1/2}}{R}, \frac{d_y^{+1/2}}{R}\right) = (-0.77, 0)$ for $\theta_{eq} = 0$. Here and in what follows, $\frac{R}{r_c} = 10^3$. The shift magnitudes continue to decrease as the ratio $\frac{\pi K}{4WR}$ increases. For example, at $\frac{\pi K}{4WR} = 5$, the shifts are only $\frac{d_x^{+1/2}}{R} = \pm 0.13$.

The direction of shifts is towards the homeotropic alignment of the director in the case of $\theta_{eq} = \pi/2$ and towards the tangential alignment for $\theta_{eq} = 0$. These directions are easy to understand: Consider an island centered at $(x, y) = (0,0)$. At the intersection of the line $x = 0$ and the NI interface, the director tilt is $45°$. The interface at $x > 0$ features the director tilt $\theta < 45°$ and the half at $x < 0$ has $\theta > 45°$. When $\theta_{eq} = 0$, a shift towards a negative $x$ increases the area of the NI interface at which $\theta < 45°$. A similar consideration explains why the tangentially anchored droplet should move to the positive end of the x-axis, Fig.5.

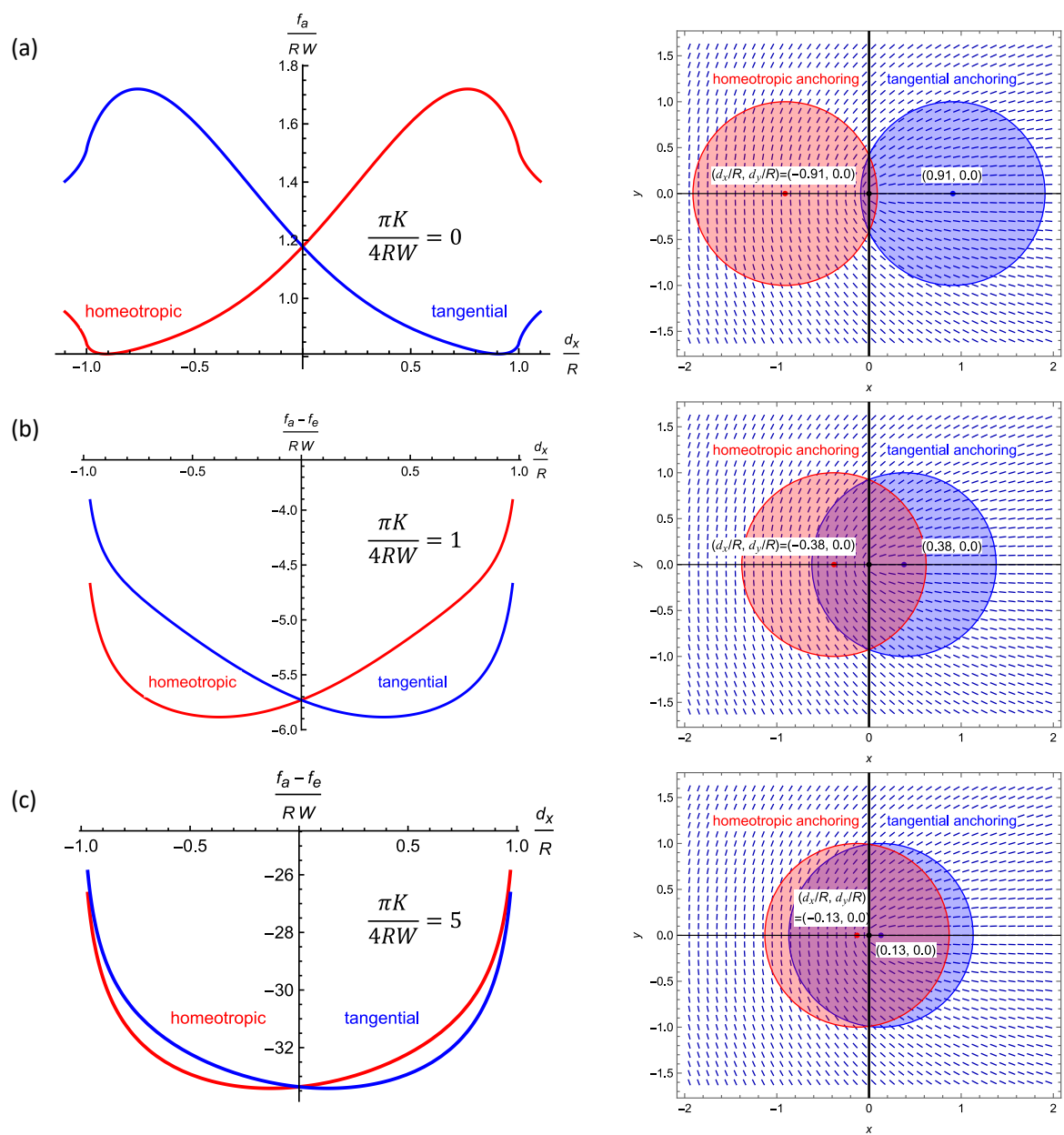

**Figure 5. Isotropic droplet with homeotropic and tangential anchoring at a +1/2 core; 2D model:** (a) anchoring effect causes a strong shift, towards the positive end of the *x*-axis, when $\theta_{eq} = 90°$ and the negative end when $\theta_{eq} = 0$; elastic contribution, (b) $\frac{\pi K}{4WR} = 1$ and (c) $\frac{\pi K}{4WR} = 5$ reduce the shifts. The right column shows the equilibrium locations of the droplets.

(iii) *In the case of conically degenerate anchoring* with $\theta_{eq} = 45°$, the isotropic islands shift along the $y$-axis, upward or downward, as follows from the minimization of $\frac{f_a}{RW}$, Fig.6a,b. There is no shift along the $x$-axis; $\left(\frac{d_x^{+1/2}}{R}, \frac{d_y^{+1/2}}{R}\right) = (0.0, \pm 0.71)$. The anchoring energy gain is not as strong as in the case of the homeotropic and tangential anchoring, and the shifts vanish when the elastic/anchoring ratio $\frac{\pi K}{4WR}$ exceeds 0.79, Fig.6c. In Fig.6c, the plots are shifted so that the energy minima are always at 0.

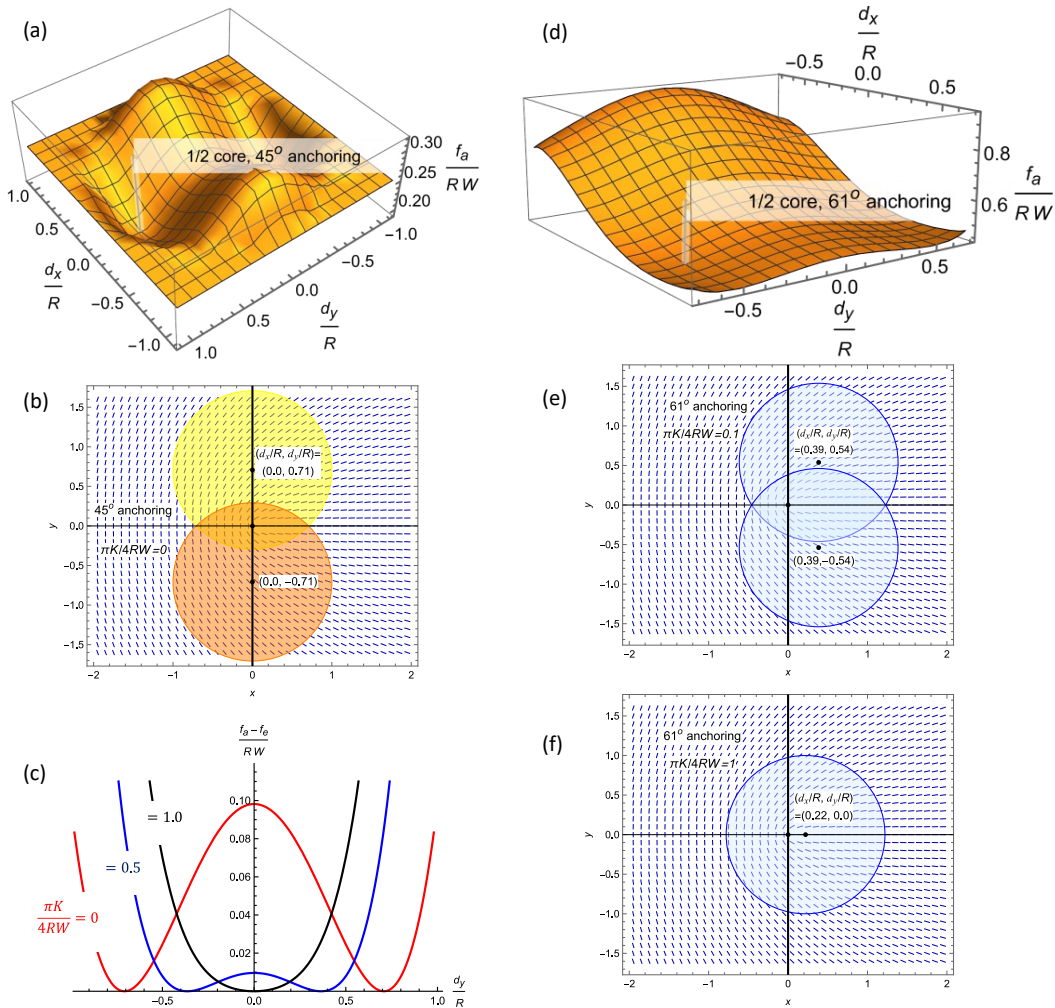

**Figure 6. Isotropic droplets with tilted anchoring at +1/2 core; 2D model:** (a,b,c) $\theta_{eq} = 45°$; the shifts are only along the $y$-axis; (d,e,f) $\theta_{eq} = 61°$; the shifts are along both $x$- and $y$-axes; (a,d) anchoring energies vs. the $x, y$ shifts for $\theta_{eq} = 45°$ and $\theta_{eq} = 61°$, respectively; (b,e) equilibrium locations of the droplets in the absence of the elastic restoring force; (c,f) elastic energy reduces the shifts along the $y$-axis for $\theta_{eq} = 45°$ and $\theta_{eq} = 61°$, respectively.

(iv) *Tilted conical anchoring* at $\theta_{eq} = 61°$ produces both horizontal and vertical shifts, Fig.6(d,e), with the minima of $\frac{(f_a - f_e)}{RW} = -0.17$ at $\left(\frac{d_x^{+1/2}}{R}, \frac{d_y^{+1/2}}{R}\right) = (0.39, \pm 0.54)$ when $\frac{\pi K}{4WR} = 0.1$. Note that the shift along the $x$-axis of symmetry is towards the homeotropic alignment of the director, similarly to the case of the tangential anchoring in Fig.5. As the elastic/anchoring ratio exceeds 0.65, the shifts along the $y$-axis disappear, and the I-islands moves only along the $x$-axis, Fig.6f.

The predicted shifts of the isotropic islands from the cores of the +1/2 topological defects remain qualitatively the same and quantitatively very close to the ones above if the surface anchoring potential in Eq. (2) is replaced by the "generalized" Rapini-Papoular potential,

$$\sigma_{NI} = \sigma_0 + W_{RP}\sin^2(\theta - \theta_{eq}), \qquad (6)$$

where the anchoring coefficient is $W_{RP} = (1.2 \pm 0.3) \times 10^{-6}$ J/m$^2$; the latter estimate is obtained by converting the units J/(m$^2$ rad$^2$) used in Ref. [41] to J/m$^2$. For example, the minimum of $(f_a - f_e)/RW$ remains at $\left(\frac{d_x^{+1/2}}{R}, \frac{d_y^{+1/2}}{R}\right) = (0.77, 0)$ for the tangential anchoring and at $(-0.77, 0)$ for the homeotropic anchoring, i.e., at the same locations as in the case of the potential given by Eq. (2).

According to Figs. 5 and 6, the surface anchoring at the NI interface does indeed shift the inclusions. This prediction should remain valid for any disc-like inclusion, provided the surface anchoring is strong enough compared to the elastic energy. The directions of anchoring-mediated shifts are deterministically defined by the equilibrium value of the easy angle $\theta_{eq}$. However, the directions of the anchoring-mediated shifts predicted by the model are opposite to the ones observed experimentally. For example, the model with $\theta_{eq} = 61°$ in Fig.6d,e,f predicts that the I island would shift predominantly towards the positive end of the $x$-axis, $d_x^{+1/2} > 0$, while the experiments, Fig.2,3,4, show a directly opposite polarity, $d_x^{+1/2} < 0$. A similar disagreement is found for the -1/2 cores below.

## B. Isolated -1/2 disclination

The -1/2 disclinations are of trefoil D$_{3h}$ symmetry. An immediate consequence is that there are three equivalent directions of shifts and three minima of the anchoring energy associated with an isotropic island at an isolated -1/2 disclination core. Interestingly, the minimum of the anchoring energy alone is achieved for $d_x^{-1/2} > R$, where the I droplet is not touching the -1/2 core,

The predicted gains in energy caused by the shifted equilibrium locations of the droplets at the -1/2 cores are much smaller than in the +1/2 core case; compare Fig.7 to Fig. 5,6. Also, the directions of the shifts are opposite to what is predicted for the +1/2 cores. Most importantly, these directions are opposite to the ones observed experimentally, Figs.3,4. For example, the tangentially anchored droplet is predicted to have $d_x^{-1/2} < 0$, while the experiments with a similar $\theta_{eq} = 61°$ anchoring yield $d_x^{-1/2} > 0$.

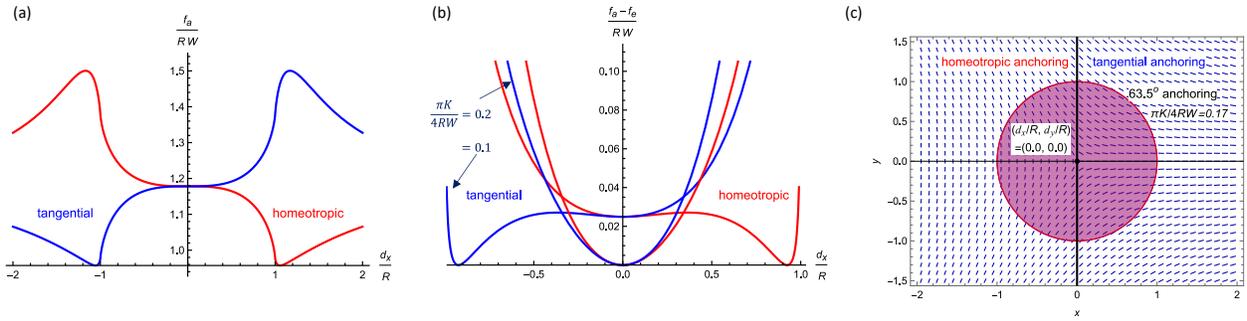

**Figure 7. Isotropic nucleus at a -1/2 core; 2D model:** (a) anchoring energy $f_a/RW$ vs $x$-shifts for tangential and homeotropic anchoring; (b) sum of anchoring and elastic energy $(f_a - f_e)/RW$ for homeotropic and tangential anchoring vs $x$-shifts; (c) an I droplet remains pinned at the core of a -1/2 defects for $\frac{\pi K}{4WR} \geq 0.17$, regardless of the easy axis direction.

The 2D model does not account for the curved shape of the NI interface observed in Fig. 12 in Appendix A . In Appendix B, we present a 3D numerical model with a curved NI interface in order to establish whether this curvature may be responsible for the difference between the predicted and observed isotropic island shifts. Numerical studies are based on the Landau-de Gennes free energy. The isotropic islands are considered as concave surfaces of revolution. The surface anchoring is degenerate conical with $\theta_{eq} = 61°$, as expected for E7. When the elastic constants of bend $K_{33}$ and splay $K_{11}$ are assumed to be equal to each other, the 3D anchoring model predicts an insignificant shift from the -1/2 cores along the positive direction of the $x$-axis, which is consistent with the experiments, Figs.3,4. However, this 3D anchoring model with $K_{11} = K_{33}$ predicts a positive, larger shift also for the +1/2 cores, which is directly opposite to the experimental observations, Figs.2-4. Below, we demonstrate that 2D and 3D models that account for the *different* values of the splay and bend elastic constants predict the shift directions that agree with the experiments.

# V. MODEL WITH DIFFERENT BEND AND SPLAY ELASTIC CONSTANTS

The Frank elastic constants are of the dimension of a force and thus can be represented as the ratio of some energy $U$ to a characteristic length $l$. In a conventional uniaxial nematic formed by rod-like molecules, the latter can only be the molecular length, $l \sim 1$ nm. The energy $U$, as suggested by P.G. de Gennes [2], should be on the order of $k_B T_c$, where $k_B$ is the Boltzmann constant and $T_c$ is the clearing temperature at which the nematic transitions into an isotropic fluid. For $T_c \sim 300$ K, one finds $K_{ii} \sim \frac{U}{l} \sim 4$ pN, which is close to the experimentally measured values. For example, elastic constants of pentylcyanobiphenyl (5CB) are listed [46] at 31°C (about 4 °C below $T_c$) as $K_{11} = 4.5$ pN for splay, $K_{22} = 3$ pN for twist, and $K_{33} = 5.5$ pN for bend. The constants follow the trend $K_{33} > K_{11} > K_{22}$ also in other nematics formed by rod-like molecules [47-50], including the E7 mixture. In particular, in E7 at 23 °C, $K_{11} = 10.8$ pN, $K_{22} = 6.8$ pN, and $K_{33} = 17.5$ pN [51]. The inequalities $K_{33} > K_{11} > K_{22}$ are preserved in the entire homogeneous nematic range of E7, although the disparity diminishes as the temperature approaches $T_c$ [52]. Despite the fact that there are no direct measurements of the elastic constants in the NI biphasic region, it is safe to assume that the N component would feature the same trend $K_{33} > K_{11} > K_{22}$. Below we discuss the effect of this disparity on the shift of the I nuclei from the disclination cores, first for 2D and then for 3D geometry.

### A. Two-dimensional model

The Frank-Oseen free energy density in two dimensions is comprised of the splay and bend terms, $\frac{1}{2}[K_{11}(\text{div }\hat{\mathbf{n}})^2 + K_{33}(\hat{\mathbf{n}} \times \text{curl }\hat{\mathbf{n}})^2]$. In the 2D model, the I island is a disk of a radius $R$ that is shifted by a distance $d < R$ from the disclination core at $(x, y) = (0,0)$ in a direction at an angle $\alpha$ away from the $x$-axis. In cylindrical coordinates $(\beta, \rho)$, with $\beta$ the polar angle and $\rho$ the radial coordinate, the boundary of the disk is given by $\rho \equiv \rho(\beta) = d\cos(\beta - \alpha) + d\sqrt{\cos^2(\beta - \alpha) - 1 + \frac{R^2}{d^2}}$. The elastic energy gain for placing the I disk at the +1/2 core has contributions from the sum and difference of the two elastic constants:

$$f_{e+} = \frac{K}{8}\int_0^{2\pi} d\beta' \int_{r_c}^{\rho(\beta')} \frac{d\rho'}{\rho'} - \frac{(K_{33}-K_{11})}{16}\int_0^{2\pi} \cos\beta'\, d\beta' \int_{r_c}^{\rho(\beta')} \frac{d\rho'}{\rho'}, \qquad (7)$$

where $K = (K_{11} + K_{33})/2$. The shift vector in Cartesian coordinates is $\mathbf{d} = \{d\cos\alpha, d\sin\alpha\}$ and $\alpha = 0$ corresponds to the shift along the positive $x$-axis directed from the head to the tail of the +1/2 defect. For the -1/2 defect, the expression is similar:

$$f_{e-} = \frac{K}{8}\int_0^{2\pi} d\beta' \int_{r_c}^{\rho(\beta')} \frac{d\rho'}{\rho'} - \frac{(K_{33}-K_{11})}{16}\int_0^{2\pi} \cos\beta'\,[2\cos(2\beta') - 1]\, d\beta' \int_{r_c}^{\rho(\beta')} \frac{d\rho'}{\rho'}. \qquad (8)$$

Note that the first terms in Eqs. (7,8), proportional to $K$, are identical to the one-constant elastic energy. Thus, the second terms in Eqs. (7,8), proportional to $K_{33} - K_{11}$, are the elastic contributions that arise due to elastic constant disparity.

The integrals over the polar angle cannot be calculated analytically for an arbitrary shift angle $\alpha$. However, we are mostly interested in the values $\alpha = 0, \pi$, i.e., the shifts $d_x$ along the $x$-axis. For these values, one finds for +1/2 disclinations

$$f_{e+} = \frac{\pi}{4}K\ln\frac{\sqrt{R^2 - d_x^2}}{r_c} - \frac{(K_{33}-K_{11})R}{4d_x\sqrt{1-(d_x/R)^2}}\left[\mathcal{K}\left(\frac{d_x^2}{d_x^2 - R^2}\right) + \left(\frac{d_x^2}{R^2} - 1\right)\mathcal{E}\left(\frac{d_x^2}{d_x^2 - R^2}\right)\right], \qquad (9)$$

where $\mathcal{K}(\ldots)$ and $\mathcal{E}(\ldots)$ are the complete elliptic integrals of the first and second kind, respectively. For -1/2 disclinations, we find

$$f_{e-} = \frac{\pi}{4}K\ln\frac{\sqrt{R^2 - d_x^2}}{r_c} - \frac{(K_{33}-K_{11})R^3}{36d_x^3\sqrt{1-(d_x/R)^2}}\left[\left(\frac{5d_x^2}{R^2} - 8\right)\mathcal{K}\left(\frac{d_x^2}{d_x^2 - R^2}\right) + \left(\frac{d_x^4}{R^4} - \frac{9d_x^2}{R^2} + 8\right)\mathcal{E}\left(\frac{d_x^2}{d_x^2 - R^2}\right)\right]. \qquad (10)$$

We drop the superscript +1/2 and -1/2 in $d_x$ to simplify the notations.

The balance of the anchoring and elastic energies is evaluated for the fixed difference of the bend and splay elastic constants: $K_{33} - K_{11} = 2K$. The calculation demonstrates that the disparity of $K_{11}$ and $K_{33}$ can shift I droplets in directions opposite to those caused by the surface anchoring. Namely, the shift of a nucleus with tangential anchoring at a +1/2 core reverses its sign from $d_x^{+1/2} > 0$ to the experimentally observed $d_x^{+1/2} < 0$ when the elastic to anchoring energies ratio increases to $\frac{\pi K}{4WR} > 1.58$, Fig.8a. For the experimentally relevant $\theta_{eq} = 61°$ anchoring, the critical value of $\frac{\pi K}{4WR}$ causing a sign reversal of $d_x^{+1/2}$ is even smaller, 0.84, Fig.8b. An inclusion with homeotropic anchoring shifts to the negative end of the $x$-axis since both the anchoring and the anisotropic elastic energy favor such a shift, Fig.8c. Note that an increase of $\frac{\pi K}{4WR}$ in this case makes the shifts smaller, as the overall effect of an "isotropic" elastic energy is to resist the shifts from the strongly deformed core.

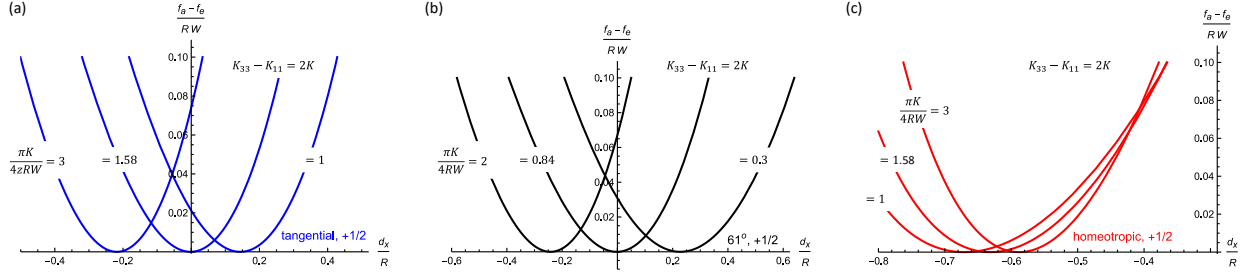

**Figure 8. Isotropic islands at a +1/2 core for $K_{33} - K_{11} = 2K$, 2D model:** (a) shift $d_x^{+1/2}$ of a tangentially anchored droplet changes the sign from negative to positive as the ratio $\frac{\pi K}{4WR}$ increases past 1.58; (b) for $\theta_{eq} = 61°$ anchoring, the critical value of $\frac{\pi K}{4WR}$ causing a sign reversal of $d_x^{+1/2}$ is 0.84; (c) homeotropically anchored island shifts to the negative end of the $x$-axis; an increase of the isotropic elastic energy reduces the magnitude of shifts.

The case of the -1/2 cores is similarly controlled by the balance of the anchoring and elastic energies, Fig.9. However, in this case, the energy vs. shift dependencies are very shallow, as compared to the +1/2 case, Fig.9. It implies that the shifts can be affected strongly by elastic interactions between the droplets at oppositely charged disclinations and by the meniscus effects. The meniscus effect would realign the director parallel to the latitude lines of the catenoid-like NI interface (see Appendix B) to reduce the elastic distortions energy through the so-called geometrical anchoring [53]. All these effects can be accounted for only by elaborate numerical analysis. We made an initial step in this direction by showing that the 3D model with $K_{33} > K_{11}$, described in detail in Appendix B, predicts the same polarities of shifts as in the experiment.

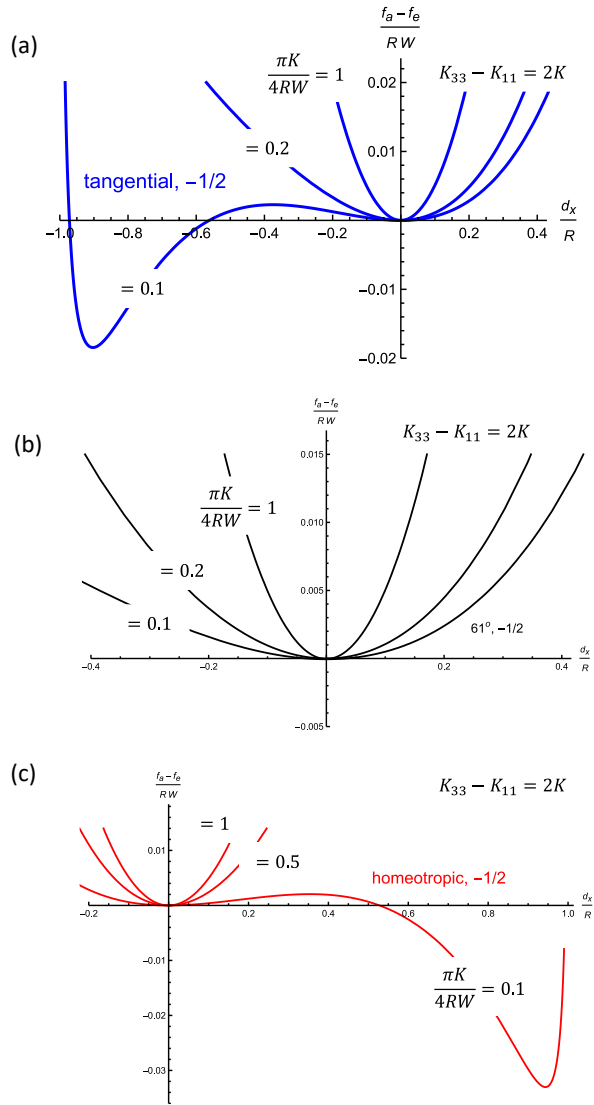

**Figure 9. Isotropic islands at a -1/2 core for $K_{33} - K_{11} = 2K$, 2D model:** elastic energy reduces the shifts $d_x^{-1/2}$ of I droplets for (a) tangential, (b) tilted, $\theta_{eq} = 61°$, (c) homeotropic anchored droplets.

## B. Three-dimensional model

As discussed in Appendix B (see Fig. 13), a 3D model with $K_{11} = K_{33}$ shows that the meniscus effect does not explain the shift of the I islands at a +1/2 defect. The 2D analysis presented above shows that the inequality $K_{33} > K_{11}$ is required to reconcile the theory and the experiment. The 3D model corroborates this finding when the disparity of the elastic constants is accounted for. Figure 10(a) shows that the I islands do indeed prefer to move in the directions observed in the experiment when we take typical elastic constant values: $K_1 \approx 2.0$ pN, $K_3 \approx 3.0$ pN, and $K_2 \approx 1.3$ pN. The director fields surrounding the isotropic islands at their optimal locations are shown in Fig. 10(b,c).

We see in Fig. 10 that the I nucleus at the +1/2 defect now moves in the *opposite* direction compared to the equal constants $K_{11} = K_{33}$ prediction and has a significant shift toward bend (tangentially anchored) region, with an optimal displacement $d_{\min}/R \approx -0.96$, consistent with the experimental results and the 2D analysis. Conversely, an island at the -1/2 defect exhibits a much smaller shift toward the *homeotropically* anchored region, with an optimal displacement of $d_{\min}/R \approx 0.11$, again in line with the experimental observations, Figs.2-4 Interestingly, this shift is somewhat larger than the one predicted by the two-constant 3D model with $K_{11} = K_{33} > K_{22}$, see Fig. 13 in Appendix B.

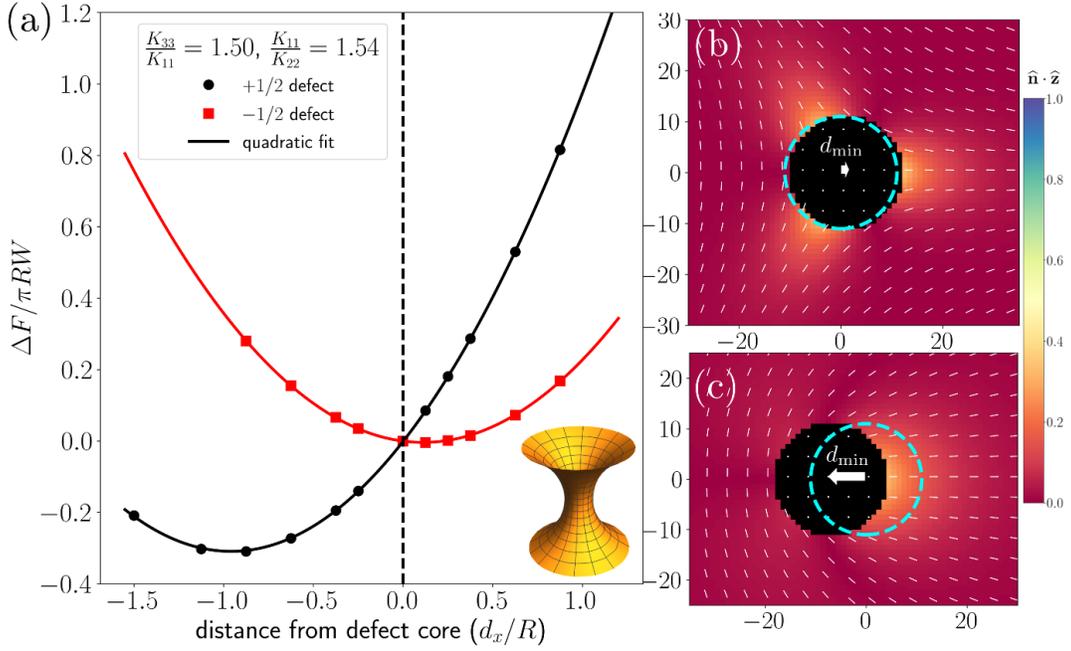

**Figure 10. Three-dimensional simulation of an isotropic droplet with unequal splay and bend.** (a) The free energy difference between an isotropic island placed at the center of a $\pm\frac{1}{2}$ core versus one at a displacement $\frac{d_x}{R}$ away ($d_x > 0$ toward the homeotropic region). The island is modeled as a catenoid of revolution, shown in the inset, with a radius $R$ at the midplane and a degenerate conical anchoring ($\theta_{eq} = 61°$). Quadratic fits to the free energy determine the preferred isotropic island position: $d_{\min}/R = 0.11$ for the $-1/2$ defect in (b) and $\frac{d_{\min}}{R} = -0.96$ for the $+1/2$ defect in (c). Horizontal slices through the cell showing the director fields for the $-1/2$ (b) and $+1/2$ (c) defects at their respective free energy minima, with the color representing the vertical component. The $-1/2$ is very slightly shifted toward the homeotropic region, while the $+1/2$ has a more dramatic shift

to the tangential anchoring side, consistent with the experimental results. The slices are taken in the same vertical position as in Fig. 13(b,c) in Appendix B.

## VI. CONCLUSION

We considered a model system of singular $\pm 1/2$ disclinations interacting with a disk-like foreign particle. The experimental realization of the foreign inclusion is an isotropic nucleus, which is stable in the biphasic NI region. The experiments show that the inclusion does not stay centered at the disclination core at which it nucleates. Instead, as the temperature increases, the I droplet shifts from the core. The shift is larger in the case of the +1/2 disclinations. Our analysis shows that the shifts can be caused by numerous mechanisms, such as surface anchoring, elastic interactions of the two droplets located at the cores of oppositely charged disclinations, and elastic interactions of the droplets with the deformed director field outside the droplet. Although the surface anchoring explains the shifts, the predicted direction is opposite to what is observed experimentally. The shift observed in experiments is predicted by a model that accounts for the difference between the bend and splay elastic constants. When $K_{33} > K_{11}$, the islands move towards the head of a +1/2 disclination, where the director bends; replacing the strongly bent nematic with an isotropic island reduces the bend elastic energy. The finding is important from the fundamental point of view since one often resolves to the so-called one-constant approximation in the description of liquid crystals, which would produce a wrong result for E7. Usually, the elastic constant disparity is considered to be relevant in the deep nematic phase or in non-conventional phases such as ferroelectric and twist-bend nematic [54], but the present study shows that it might be relevant even in the proximity of the clearing temperature and in the biphasic region. We also conducted experiments with a nematic CCN-47, in which $K_{11}$ and $K_{33}$ are very close to each other even far away from the clearing point [55]; in this case, the shifts were insignificant and hard to quantify.

The analysis above should be valid for other systems in which nematic disclinations control assemblies of colloidal particles, a subject of significant interest [4,7-9,13,16,28,29,31,56-64]. The effect of the elastic constants disparity would be even stronger in the deep nematic phase, well below the clearing temperature. It is important to stress that the balance of the surface anchoring $\propto WR$ and elastic $\propto K\ln R/r_c$ terms changes with the size $R$ of the inclusion. For small $R$, the positions are defined mostly by the elastic contributions, while for large inclusion, the locations are dictated by the surface anchoring. In other words, the amplitude and even the polarity of shifts can vary with the size of the inclusion. The transition between these regimes should occur when $R \approx (K/W)\ln R/r_c$. Our preliminary experiments show that, in line with these expectations, the shift $|d_x^{+1/2}|$ of the I nuclei from the +1/2 defect cores first increases with temperature for small radii $R < (K/W)\ln R/r_c$, and then decreases as the island radius increases to $R > (K/W)\ln R/r_c$. However, the shape of islands larger than $R \approx 25$ μm is no longer disk-like, which makes the analysis of this non-monotonous dependence difficult. As already stated, a complete description of the behavior of the inclusions at the defect cores should consider several important factors, such as the 3D geometry of the interfaces, variation of the scalar order parameter, shape changes, interactions between disclinations, etc. Our analysis is limited to the simplest cases in which the analytical and numerical calculations are sufficiently transparent to see the underlying physical mechanisms. The demonstrated dependence of inclusions' locations on the peculiarities of the surface anchoring potential and the anisotropy of elastic constants suggests a pathway to a rational design of disclination-assisted microscale assemblies.

# VIII. ACKNOWLEDGMENTS

We thank Qi-Huo Wei and Sergij Shiyanovskii for fruitful discussions. ODL acknowledges the support of the NSF grant DMS-2106675. Computational support was provided by the Department of Earth, Environment, and Physics at Worcester State University.

# APPENDIX A. ANCHORING AND WETTING GEOMETRY DETAILS

To estimate the angle $\theta_{eq}$ between the director and the normal to the E7 NI interface, we use glass plates with spin-coated polystyrene (PS), 0.5 wt% solution in chloroform. PS sets a degenerate in-plane orientation of cyanobiphenyls [42]. A small temperature gradient is created by inserting a 20 μm thick insulator between the cell and the hot stage at one edge of the cell. The temperature gradient results in an extended NI interface. The PolScope MicroImager's map of the director, Fig.11, suggests that $\hat{n}_{eq}$ is tilted by $\theta_{eq} = (61 \pm 8)°$ from the normal to the interface.

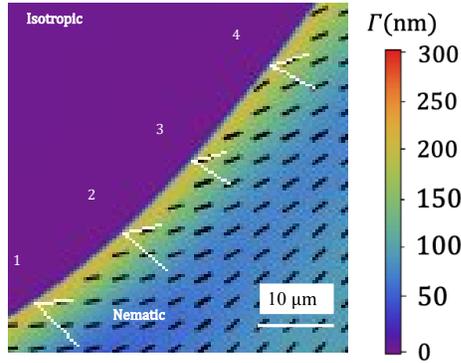

**Figure 11. Anchoring of E7 director at the NI interface**. PolScope MicroImager texture of a thin $(1.8 \pm 0.1)$ μm slab of E7 in a temperature gradient. The nematic director makes an angle $\theta_{eq} = (61 \pm 8)°$ with the normal to the NI interface. The optical retardance of the N phase is in the second order thus the blue pseudocolors correspond to retardance higher than 300 nm.

To establish the wetting preference, we placed a circular glass capillary with E7 in a temperature gradient, Fig.12. Polarizing microscopy confirms the meniscus effect, as the I phase wets the glass wall better than the N phase. In the cell's midplane, the NI interface's tip protrudes by about $h/3$ into the I region. To reduce the meniscus effect, the experiments on isotropic droplets nucleating at the disclination cores were performed for thin cells of thickness $h = (1.2 - 3.3)$ μm, so that the meniscus elevation is about 1 μm or less, smaller than the radius $R$ of the droplets, which allows us to treat the problem in 2D. Nevertheless, a 3D model is also considered in the analysis of the anchoring mechanism (see Appendix B and main text).

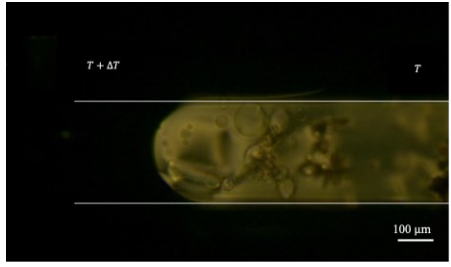

**Figure 12. NI meniscus of E7 in a capillary.** Polarized optical microscope texture of a cylindrical capillary with E7 in a temperature gradient. The isotropic phase wets the surface better than the N phase.

# APPENDIX B. 3D NUMERICAL MODELLING

To investigate the effects of the three-dimensional geometry of the NI interface, we minimized the Landau-de Gennes free energy of the nematic in a rectangular cell with a single I droplet modeled as a catenary of revolution. The isotropic islands are considered as concave surfaces of revolution, with the $z$-axis being the axis of rotational symmetry. The isotropic island is placed at the core of the disclination; the surface anchoring is taken to be degenerate tangential for simplicity. The pre-imposed director field of the +1/2 disclination is the same as in Fig. 5, i.e., it tends to be tangential to the left side of the NI interface (where the "head" of the defect is located) and perpendicular to the right-hand side of the interface (where the "tail" is located).

Specifically, we consider a symmetric, traceless 3x3 matrix $Q_{\alpha,\beta}(\mathbf{x})$ defined on each site $\mathbf{x}$ of a cubic lattice with lattice spacing $\Delta x$. The thermodynamic potential portion of the free energy density on the lattice, in the bulk of the cell, is given by

$$F_b = \frac{A}{2} Q_{\alpha\beta} Q_{\beta\alpha} + \frac{B}{2} Q_{\alpha\beta} Q_{\beta\alpha} Q_{\gamma\alpha} + \frac{C}{2} \left( Q_{\alpha\beta} Q_{\beta\alpha} \right)^2 \quad (1)$$

where $A, B, C$ are constants and summation over repeated indices is implied. These terms come from thermodynamic considerations and yield the scalar order parameter $S_0 = (-B + \sqrt{B^2 - 24AC})/6C$. We choose free energy units such that $A = -1$, and set the other parameters to $B = -12.33$ and $C = 10.06$, which yields $S_0 = 0.533$, a value consistent with the nematic order of 5CB [65]. We checked that the choices of these parameters did not qualitatively change the simulation results. In addition to this thermodynamic contribution, we have the elastic free energy density in the two-constant approximation (equal splay and bend)

$$F_e = \frac{L_1}{2}\left(\epsilon_{\alpha\gamma\delta}\partial_\gamma Q_{\delta\beta}\right)^2 + \frac{L_2}{2}\left(\partial_\beta Q_{\beta\alpha}\right)^2, \tag{2}$$

where $\partial_\alpha$ are derivatives with respect to the $\alpha = x, y, z$ directions and $\epsilon_{\alpha\gamma\delta}$ is the Levi-Cevita symbol. The constants $L_{1,2}$ are chosen to correspond to the elastic constants of 5CB, for which $K_{11} = K_{33} \approx 2.3$ pN and $K_{22} \approx 1.3$ pN, yielding $L_1 = 0.027$ and $L_2 = 0.0658$ for a lattice spacing of $\Delta x \approx 15$ nm. At the boundaries of the cell, we minimize interfacial free energy with a density given by the Rapini-Papoular potential $F_a^{(1)} = W_0\left(Q_{\alpha\beta} - Q_{\alpha\beta}^i\right)^2$ [66], where $Q_{\alpha\beta}^i = \frac{3}{2}S_0\left(\nu_\alpha\nu_\beta - \frac{1}{3}\delta_{\alpha\beta}\right)$ is the tensor corresponding to a unit vector field $\nu_\alpha \equiv \nu_\alpha(\mathbf{x})$ describing either $\pm 1/2$ defect: $\nu_\alpha = (\cos\phi_\pm, \sin\phi_\pm, 0)$ with $\phi_\pm(\mathbf{x}) = \pm 0.5 \tan^{-1}(y/x)$. We choose $W_0 = 1.0$, which corresponds to an anchoring strength of $6.6 \times 10^{-3}$ J/m$^2$, ensuring that the director follows the patterned template on the top and bottom of the cell.

The I droplet, modelled as a solid object with certain anchoring conditions, is placed in the center of the cell as a catenary of revolution spanning the cell along the $z$-axis. Specifically, the boundary of the island is given by $(\rho\cos\phi, \rho\sin\phi, z)$, where $\rho = \rho_0 \cosh(z/L)$ and we vary $z$ over the cell thickness, $-\frac{h}{2} < z < \frac{h}{2}$, and $\phi$ between 0 and $2\pi$. The parameters $\rho_0$ and $L/h$ control the width and aspect ratio of the catenary, respectively. We use a cell thickness $t = 64\Delta x \approx 1$ μm, I droplet radius $\rho_0 = 8\Delta x$, and $L/h = 1/4$. The area of the simulated cell was $128\Delta x \times 128\Delta x \approx 4$ μm$^2$. The profile of the isotropic island is shown in the right panels of Fig. 13(b,c). We implement degenerate conical anchoring conditions at an equilibrium angle $\theta_e$ (between the director and surface normal) on the catenary NI interface using a modification of the Fournier-Galatola potential [67] described in Ref.[68], with a free energy density given by $F_a = W_1\left[P_{\alpha\gamma}\widetilde{Q}_{\gamma\delta}P_{\delta\beta} - \frac{3}{2}S_0 P_{\alpha\beta}\cos^2\theta_e\right]^2$, where $\widetilde{Q}_{\alpha\beta} = Q_{\alpha\beta} + \frac{1}{2}S_0\delta_{\alpha\beta}$ and $P_{\alpha\beta} = \nu_\alpha\nu_\beta$ is a projection tensor (the vector $\nu_\alpha$ here corresponds to the local surface normal). We choose an equilibrium angle $\theta_e = 61°$ and $W_1 = 0.01$, which corresponds to an anchoring strength of $3.3 \times 10^{-5}$ J/m$^2$, yielding a dimensionless ratio $\frac{\pi K}{4W_1 R} \approx 0.4$. When the scalar order parameter does not vary substantially in the nematic region, then this anchoring energy is equivalent to the one discussed in the main text, Eq. (3), in the Frank-Oseen formulation.

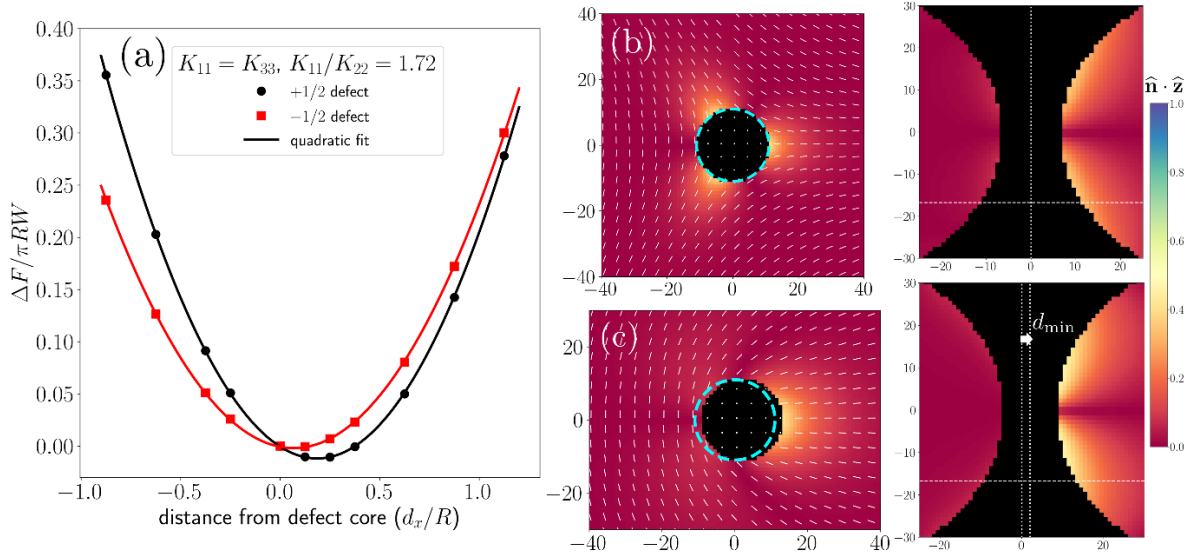

**Figure 13. Three-dimensional simulation of an isotropic droplet located at $\pm\frac{1}{2}$ defect cores.** (a) The free energy difference between an isotropic island placed at the center of a $\pm\frac{1}{2}$ core versus one at a displacement $\frac{d_x}{R}$ away ($d_x > 0$ toward the homeotropic anchoring). The island is modeled as a catenoid of revolution, shown in the inset, with a radius $R$ at the midplane and a degenerate conical anchoring ($\theta_{eq} = 61°$). Quadratic fits to the free energy determine the preferred isotropic island position: $d_{min}/R = 0.067$ for the $-1/2$ defect and $d_{min}/R = 0.19$ for the $+1/2$ defect. (b) Horizontal slice through the cell, showing the director field for the $-1/2$ defect for $d_x = 0$. The director tilts away from the plane of the cell in the homeotropic regions with unfavorable anchoring. The vertical slice in the right panel shows where we take the slice (horizontal white dashed line). We also see the profile of the catenary and its center-line (vertical white dotted line). (c) Same slice for $+1/2$ core, with the near-optimal value $d_x/R = 0.25$. The dashed circle indicates the isotropic island for $d_x = 0$. The island yields an energetically favorable configuration when shifted to the right a distance $d_{min}$ from the defect core, as indicated by the vertical dotted lines in the left panel.

The combined free energy (bulk and boundary terms) is minimized over the 128x128x64 cells of the cubic lattice using a conjugate gradient method from ALGLIB package (https://www.alglib.net). For more details on the numerical algorithm and the overall approach to minimizing this free energy, see Ref. [69]. We then vary the position of the catenary surface to explore the free energy difference between various positions of the I droplet. We find that for both the $\pm 1/2$ defects, the island tends to shift toward the homeotropic anchoring region. This makes sense as, just as in the 2D analysis, it is more favorable for the island to shift toward the side with the unfavorable anchoring. Unlike the 2D case, however, we also observe a tilting of the nematic director out of the plane of the cell. This tilt can be seen in both the cross sections in Fig. 13(b,c). Also, the shift in the +1/2 defect case is more pronounced: We estimate that the preferred position of the island is $d_{\min}/R \approx 0.19$ from the defect core, as compared to $d_{\min}/R \approx 0.067$ for the $-1/2$ defect.

The 3D model of NI interface accounts for the meniscus effect and predicts the shifts of the I nuclei from the defect cores. It shows a very small shift of the isotropic islands towards the positive end of the x-axis in the case of the $-1/2$ defect, which is close to the experimental observations in Figs. 2-4. However, it fails to reproduce even the correct direction of the experimentally observed shifts for $+1/2$ defects, Figs.2-4. The two dimensional model that accounts for the different values of the splay and bend elastic constants presented in the main text predicts the directions of shifts that agree with the experiments. In light of this, we have also checked the 3D model in the three-constant approximation.

Extending the Landau-de Gennes approach to unequal splay $K_{11}$ and bend $K_{33}$ constants introduces a much wider range of possible terms in the elastic portion of the free energy as we must introduce a cubic term in the $Q_{\alpha\beta}$ order parameter. There are many such possible terms and we adopt the choice of Ravnik and Žumer [65], where the (non-dimensional) elastic contribution reads

$$F_e = \frac{L_1}{2} \partial_\gamma Q_{\alpha\beta} \partial_\gamma Q_{\alpha\beta} + \frac{L_2}{2} \partial_\beta Q_{\alpha\beta} \partial_\gamma Q_{\alpha\gamma} + \frac{L_3}{2} Q_{\alpha\beta} \partial_\alpha Q_{\gamma\delta} \partial_\beta Q_{\gamma\delta}, \qquad (3)$$

with the term proportional to $L_3$ the higher-order cubic term. With this term, it is possible to consider unequal splay and bend. Keeping all of our other terms in the free energy the same, we set $L_1 \approx 0.03303$, $L_2 \approx 0.02831$, and $L_3 \approx 0.02530$, which corresponds to the elastic constants $K_1 \approx 2.0$ pN, $K_3 \approx 3.0$ pN, and $K_2 \approx 1.3$ pN. The results, shown in Fig. 10, agree with the experiments, Figs.2-4, and are strikingly different from the two-constant case in Fig. 13.